\documentclass[twocolumn,showpacs,prl]{revtex4}
\usepackage{graphicx}
\usepackage{dcolumn}
\usepackage{bm}
\usepackage{epsfig,psfig,pstricks,pst-grad,fancybox,graphics}
\usepackage{latexsym}
\usepackage{amsmath}
\usepackage{amssymb}
\usepackage{enumerate}
\usepackage{bbm}
\usepackage[latin1]{inputenc}

\begin{document}

\title{Superfluidity of trapped dipolar Fermi gases.}
\author{M.A. Baranov$^{1,2}$, \mbox{\L}. Dobrek$^{1}$, 
and M. Lewenstein$^{1}$}
\affiliation{(1) Institut f\"ur Theoretische Physik, Universit\"at Hannover, 
D-30167 Hannover,Germany\\
(2) Russian Research Center Kurchatov Institute, Kurchatov sq. 1, 
123183 Moscow, Russia}
\date{\today }

\begin{abstract}
We derive the phase diagram for ultracold trapped dipolar Fermi gases. Below
the critical value of the dipole-dipole interaction energy, the BCS
transition into a superfluid phase ceases to exist. The critical dipole
strength is obtained as a function of the trap aspect ratio. The order
parameter exhibits a novel behavior at the criticality.
\end{abstract}

\maketitle

The quest for the superfluid (BCS) transition \cite{deGennes} in trapped
Fermi gases is one of the most challenging goals of modern atomic, molecular
and optical physics. The possibility of the BCS transition for trapped gases
with attractive short range interactions has been predicted in \cite%
{shortrangeBCS}, and has been a subject of extensive experimental research
since then. In typical experiments evaporative cooling is used to cool
fermions. However, as the Pauli principle forbids the $s$-wave scattering
for fermions in the same internal state, Fermi-Fermi \cite{fermi-fermi} or
Fermi-Bose \cite{fermi-bose} mixtures have to be used to assure collisional
thermalization of the sample. Such a combination of evaporation and
sympathetic cooling allows to reach temperatures $T\simeq 0.1T_{F}$, where $%
T_{F}$ is the Fermi temperature at which the gas exhibits quantum
degeneracy. Unfortunately, predicted critical temperatures for the BCS
transition, $T_{c}$, are much smaller than $T_{F}$. One way to circumstance
this difficulty is to increase $T_{c}$; this way may be achieved by
increasing the strength of the atomic interactions employing a Feshbach
resonance, which allows to make the atomic scattering length $a_{s}$ large
negative. Such \textquotedblright resonance superfluidity\textquotedblright\
should lead to superfluid transition at $T_{c}\approx 0.1T_{F}$ \cite%
{resonanceSF}. Another way is to use the cooling scheme that can overcome
the effects of Pauli blocking, such as appropriately designed laser cooling 
\cite{lasercool}. Yet another promising route is to locate the ultracold gas
in an optical trap and enter the strongly correlated regime by controlling
the lattice potential \cite{corrlattice}.

The temperature of the BCS transition in a two-component Fermi gas, however,
depends also drastically on the difference of the concentrations of the two
components, which presents another experimental obstacle. This problem is
not relevant for a polarized Fermi gas with long-range interactions, such as
dipole-dipole interactions, and that is why there has been a considerable
interest recently in studying the BCS transition in dipolar Fermi gases. The
possibility of the Cooper pairing has been predicted in Refs.\cite%
{dippairing}. The critical temperature (including many-body corrections) and
the other parameters have been obtained in Ref.\cite{dipoleTc}. At this
point it is worth mentioning that possible realizations of dipolar gases
include ultracold heteronuclear molecular gases \cite{Meijer}, atomic gases
in a strong DC electric field \cite{YouMarinesku}, atomic gases with
laser-induced dipoles \cite{laserinddip}, or with magnetic dipoles \cite%
{PfauDoyle}. For dipolar moments $d$ of the order of one Debye and densities 
$n$ of $10^{12}\mathrm{cm}^{-3}$, $T_{c}$ should be in the range of $100%
\mathrm{nK} $, i.e. experimentally feasible.

Dipole-dipole interaction is not only of long-range, but also anisotropic,
i.e. partially attractive and partially repulsive. Thus for trapped gases,
the nature of the interaction may be controlled by the geometry of the trap.
For a dipolar Bose gas in a cylindrical trap with the axial (radial)
frequency $\omega _{z}(\omega _{\rho })$, there exist a critical aspect
ratio $\lambda =(\omega _{z}/\omega _{\rho })^{1/2}$, above which the
Bose-condensed gas collapses if the atom number is too large \cite{Santos},
and below which the condensate exhibits the roton-maxon instability \cite%
{Santos-roton}. The effects of trap geometry are expected to dominate also
the physics of trapped dipolar Fermi gases \cite{Nobel}. So far, we have
only been able to derive analytic corrections to $T_{c}$ in ''loose'' traps,
and to consider the case of an infinite ''slab'' ($\omega _{\rho }=0$, $%
\omega _{z}$ is finite). In this case there exists a critical frequency
above which the dipole interaction is predominantly repulsive, and the
superfluid phase ceases to exist.

In this Letter we present the ultimate solution of the fundamental problem
of the effect of trap geometry on the BCS transition in trapped dipolar
Fermi gases. We calculate the phase diagram in the plane $\Gamma
-\lambda^{-1}$, where $\Gamma \sim nd^{2}/\mu $ is the dipole-dipole
interaction energy per particle in the units of the chemical potential $\mu $%
. Below the critical value $\Gamma <\Gamma _{c}$, the BCS transition does
not take place. We determine dependence of $\Gamma _{c}$ on $\lambda $, and
calculate the order parameter at the criticality. The order parameter
exhibits a novel oscillatory behavior in strongly elongated cylindrical
traps.

We consider a dipolar \ single-component Fermi gas in a cylindrically
symmetric trap described by the Hamiltonian

\begin{eqnarray}
\widehat{H} &=&\int_{\mathbf{r}}\widehat{\psi }^{\dagger }(\mathbf{r})\left[
-\frac{\hbar ^{2}\nabla ^{2}}{2m}+V_{\mathrm{trap}}(\mathbf{r})-\mu \right] 
\widehat{\psi }(\mathbf{r})  \label{Hamiltonian} \\
&&+\frac{1}{2}\int_{\mathbf{r},\mathbf{r}^{\prime }}\widehat{\psi }^{\dagger
}(\mathbf{r})\widehat{\psi }^{\dagger }(\mathbf{r}^{\prime })V_{\mathrm{dip}%
}(\mathbf{r}-\mathbf{r}^{\prime })\widehat{\psi }(\mathbf{r}^{\prime })%
\widehat{\psi }(\mathbf{r}),  \notag
\end{eqnarray}%
where $m$ is the atoms mass, $V_{\mathrm{trap}}(\mathbf{r})=m[\omega _{\rho
}^{2}(x^{2}+y^{2})+\omega _{z}^{2}z^{2}]$ the trapping potential, $\mu $ the
chemical potential, and $V_{\mathrm{dip}}(\mathbf{r}%
)=(d^{2}/r^{3})(1-3z^{2}/r^{2})$. The dipoles are assumed to be polarized
along the $z$-direction, $\widehat{\psi }^{\dagger }(\mathbf{r})$ and $%
\widehat{\psi }(\mathbf{r})$ are the atomic creation and polarization
operators that fulfill canonical anticommutation relations. Our aim is to
apply the BCS theory (see. e.g. \cite{deGennes}) to the system described by
Eq. (\ref{Hamiltonian}), calculate the critical temperature, and the
superfluid order parameter $\Delta (\mathbf{r},\mathbf{r}^{\prime })=V_{%
\mathrm{dip}}(\mathbf{r}-\mathbf{r}^{\prime })\left\langle \widehat{\psi }(%
\mathbf{r})\widehat{\psi }(\mathbf{r}^{\prime })\right\rangle $. In
particular, we are interested in the critical value of the aspect ratio $%
\lambda $ below which the BCS pairing does not take place for a given
strength of the dipole interaction. The BCS theory leads to the gap equation
which is a linear integral equation for $\Delta (\mathbf{r},\mathbf{r}%
^{\prime })$ (a function of 6 variables) with a kernel that depends on 12
coordinates. Numerical solution of this equation is impossible without
further analytical simplifications. As shown in Ref.\cite{dipoleTc}, the
dominant contribution to the BCS pairing comes from the $p$-wave scattering
(contributions of higher angular momentum, although present due to the
long-range character of the dipole-dipole interaction, are numerically
small) with zero projection of the angular momentum on the $z$-axis, $%
l_{z}=0 $, in which the interaction is attractive (see also Ref. \cite%
{dippairing}). In the $p$-wave channels with $l_{z}=\pm 1$ the interaction
is repulsive and, therefore, results only in renormalizations of a
Fermi-liquid type. The latter are controlled by the small parameter of the
theory $\Gamma $ and will be neglected. Thuerefore, for the considered
pairing problem we can model the dipole-dipole interaction by the following
(off-shell) amplitude

\begin{equation}
\Gamma _{1}(\mathbf{p},\mathbf{p}^{\prime },E)=p_{z}p_{z}^{\prime }%
\widetilde{\gamma }_{1}(E),  \label{gammadef}
\end{equation}%
where $p$ is the incoming momentum, $p^{\prime }$ the outgoing one, and $%
\widetilde{\gamma }_{1}(E)$ some function of the energy $E$. The amplitude $%
\Gamma _{1}$describes anisotropic scattering only\ in the $p$-wave channel
with the projection of the angular momentum $l_{z}=0$. The function $%
\widetilde{\gamma }_{1}(E)$ obeys the equation

\begin{eqnarray}
\widetilde{\gamma }_{1}(E)-\widetilde{\gamma }_{1}(E^{\prime })
&=&\int^{\Lambda }\frac{d\mathbf{p}}{(2\pi )^{3}}\widetilde{\gamma }%
_{1}(E)\left\{ \frac{p_{z}^{2}}{p^{2}-E+i0}\right.  \notag \\
&&\left. -\frac{p_{z}^{2}}{p^{2}-E^{\prime }+i0}\right\} \widetilde{\gamma }%
_{1}(E^{\prime }),  \label{gammaeq}
\end{eqnarray}%
that follows from the Lipmann-Schwinger equation for the off-shell
scattering amplitude, and is assumed to be negative, $\widetilde{\gamma }%
_{1}(E)<0$, in order to guarantee the BCS pairing. The cut-off parameter $%
\Lambda $ ensures the convergence of the integral and, in fact, can be
expressed in terms of the observable scattering data corresponding to
on-shell scattering with $p=p^{\prime }$ and $E=p^{2}/m$. Tt follows from
the above equation that $\widetilde{\gamma }_{1}(E)\,$is inversely
proportional to $E$, $\widetilde{\gamma }_{1}(E)=\gamma _{1}\ (2mE)^{-1}$,
with some coefficient $\gamma _{1}$. Therefore, the on-shell amplitude is
energy independent, as it should be for low-energy scattering on the
dipole-dipole potential (see Ref. \cite{lowEdipscattering}).

The coefficient $\gamma _{1}$ determines the value of the critical
temperature $T_{c}$ of the BCS transition in a spatially homogeneous gas and
hence, using the results of Ref.\cite{dipoleTc}, can be expressed through
the dipole moment $d$. In this case the order parameter has the form $\Delta
(\mathbf{p})=p_{z}\Delta _{0}$ with some constant $\Delta _{0}$, and the
linearized gap equation results in the equation for the critical temperature 
$T_{c}$

\begin{equation}
\frac{1}{\widetilde{\gamma }_{1}(\mu )}=-\int \frac{d\mathbf{p}}{(2\pi )^{3}}%
\frac{p_{z}^{2}}{2\xi _{p}}\left[ \tanh \frac{\xi _{p}}{2T_{c}}-1\right] ,
\label{gaphom}
\end{equation}%
where $\xi _{p}=p^{2}/2m-\mu $ and the bare interaction is renormalized in
terms of the scattering amplitude with $\widetilde{\gamma }_{1}(\mu )=\gamma
_{1}/p_{F}^{2}$ at the Fermi energy $\varepsilon _{F}=\mu =p_{F}^{2}/2m$
along the lines of Ref. \cite{GorkovMelikB} ($p_{F}$ is the Fermi momentum).
After integrating over $p$ we obtain the following equation for $T_{c}$:

\begin{equation}
1=\frac{1}{3}\left\vert \gamma _{1}\right\vert \nu _{F}\left[ \ln \frac{2\mu 
}{T_{c}}-\frac{8}{3}-\ln \frac{\pi }{4}+C\right] ,  \label{Tchom}
\end{equation}%
where $\nu _{F}=mp_{F}/2\pi ^{2}$ is the density of states at the Fermi
energy and $C=0.5772$ the Euler constant, and therefore $\gamma
_{1}=-24d^{2}/\pi $ in order to reproduce the result of Ref.\cite{dipoleTc}
for $T_{c}$.

In the ordinary space, the scattering amplitude $\Gamma _{1}$is

\begin{equation}
\Gamma _{1}(\mathbf{r},\mathbf{r}^{\prime },E)=\partial _{z}\delta (\mathbf{r%
})\partial _{z^{\prime }}\delta (\mathbf{r}^{\prime })\widetilde{\gamma }%
_{1}(E),  \label{gammaspace}
\end{equation}%
where $\mathbf{r}$ and $\mathbf{r}^{\prime }$ are the relative distances
between the two incoming and two outgoing particles, respectively.
Therefore, the order parameter in a trapped gas, $\Delta (\mathbf{r}_{1},%
\mathbf{r}_{2})\sim \left\langle \psi (\mathbf{r}_{1})\psi (\mathbf{r}%
_{2})\right\rangle $, has the form $\Delta (\mathbf{r}_{1},\mathbf{r}%
_{2})=\partial _{z}\delta (\mathbf{r})\Delta _{0}(\mathbf{R})$, where $%
\mathbf{r}=\mathbf{r}_{1}-\mathbf{r}_{2}$ and $\mathbf{R}=(\mathbf{r}_{1}+%
\mathbf{r}_{2})/2$, and the corresponding equation for the critical
temperature is

\begin{eqnarray}
&&\frac{\Delta _{0}(\mathbf{R})}{\widetilde{\gamma }_{1}(\mu )}=-\int_{%
\mathbf{R}^{\prime }}\left[ \sum_{\mathbf{n}_{1},\mathbf{n}_{2}}M_{\mathbf{n}%
_{1}\mathbf{n}_{2}}(\mathbf{R})M_{\mathbf{n}_{1}\mathbf{n}_{2}}(\mathbf{R}%
^{\prime })\right.  \notag \\
&&\left. \times \frac{\tanh \left( \xi _{1}/2T\right) +\tanh \left( \xi
_{2}/2T\right) }{2(\xi _{1}+\xi _{2})}-\int \frac{d\mathbf{p}}{(2\pi )^{3}}%
\int \frac{d\mathbf{q}}{(2\pi )^{3}}\right.  \notag \\
&&\left. \times \frac{p_{z}^{2}}{2\xi _{p}+q^{2}/4m}\exp (i\mathbf{q}(%
\mathbf{R}-\mathbf{R}^{\prime }))\right] \Delta _{0}(\mathbf{R}^{\prime }),
\label{gapeqren}
\end{eqnarray}%
where $\xi _{i}=\xi (\mathbf{n}_{i})$, $\mathbf{n}=(n_{x},n_{z},n_{z})$ are
the harmonic oscillator quantum numbers, $\xi (\mathbf{n})=\hbar \left[
\omega _{z}(n_{z}+1/2)+\omega _{\rho }(n_{x}+n_{y}+1)\right] -\mu $, and $M_{%
\mathbf{n}_{1}\mathbf{n}_{2}}(\mathbf{R})\equiv
M_{n_{1z}n_{2z}}^{(z)}(z)M_{n_{1x}n_{2x}}^{(\rho
)}(x)M_{n_{1y}n_{2y}}^{(\rho )}(y)$ with $M_{n_{1}n_{2}}^{(z)}(z)=\frac{1}{2}%
\left[ \varphi _{n_{1}}(z)\partial _{z}\varphi _{n_{2}}(z)-\varphi
_{n_{2}}(z)\partial _{z}\varphi _{n_{1}}(z)\right] $, $M_{n_{1}n_{2}}^{(\rho
)}(x)=\varphi _{n_{1}}(x)\varphi _{n_{2}}(x)$, and similar for $%
M_{n_{1}n_{2}}^{(\rho )}(y)$, $\varphi _{n}$ are the harmonic oscillator
wave functions.

The gap equation (\ref{gapeqren}) is still hardly tractable numerically. We
employ thus the series of additional approximations. We assume a large
number of particles in the trap and hence, the chemical potential $\mu $ is
much larger than the trap frequencies, $\mu \gg \omega _{z},\omega _{\rho }$%
. Then, for the most important for the BCS pairing states near the Fermi
energy $\varepsilon _{F}=\mu $ we can use the WKB approximation for the wave
functions $\varphi _{n}$ while calculating the functions $M_{\mathbf{n}_{1}%
\mathbf{n}_{2}}(\mathbf{R})$. Another simplification is due to the fact that
the BCS order parameter $\Delta _{0}(\mathbf{R})$ varies slowly on an
interparticle distance scale $n^{-1/3}\sim \hbar /p_{F}$, where $p_{F}=\sqrt{%
2m\mu }$ is the Fermi momentum in the center of the trap in the Thomas-Fermi
approximation. As a result, the pairing correlations are pronounced only
between close in energy states. Therefore, one can neglect $q^{2}/4m$ in the
denominator of the second term in Eq.(\ref{gapeqren}), as well as rapidly
oscillating contributions in the functions $M_{\mathbf{n}_{1}\mathbf{n}_{2}}(%
\mathbf{R})$. Under these assumptions these functions become proportional to
the Chebyshev polynomials $\mathrm{U}_{n-1}(z/l_{zN})$, $\mathrm{T}%
_{n}(x/l_{\rho N})$ and $\mathrm{T}_{n}(y/l_{\rho N})$, where $l_{iN}=\sqrt{%
2N\hbar /n\omega _{i}}=l_{0i}\sqrt{2N}$.

The critical aspect ratio $\lambda _{c}$ corresponds to vanishing critical
temperature. We therefore calculate the gap equation in the limit $T\ll
\omega _{i}$. After very tedious, but fully analytical calculations with the
substitution $\Delta _{0}(\mathbf{R})\rightarrow \Delta _{0}(\mathbf{r}%
)=\Delta _{0}(zR_{TF}^{(z)},xR_{TF}^{(\rho )},yR_{TF}^{(\rho )})$, where $x$%
, $y$, $z$ are now dimensionless $\left\vert x\right\vert $, $\left\vert
y\right\vert $, $\left\vert z\right\vert \leq 1$, and $R_{TF}^{(i)}$ denote
the radius of the gas cloud in the $i$-direction, calculated in \ the
Thomas-Fermi approximation, $R_{TF}^{(i)}=p_{F}/m\omega _{i}$, we arrive at
the gap equation

\begin{eqnarray}
&&\frac{3}{\Gamma }(1-r^{2})\Delta _{0}(\mathbf{r})=(1-r^{2})^{3/2}\left\{
\ln \frac{2\mu (\mathbf{r})}{\omega _{z}}\right. \\
&&\left. -\frac{2}{3}(4-\ln 4)\right\} \Delta _{0}(\mathbf{r})-\frac{3\pi
^{2}}{2}\int_{\mathbf{r}}^{\prime }K(\mathbf{r},\mathbf{r}^{\prime })\Delta
_{0}(\mathbf{r}^{\prime }),  \notag  \label{gapeqfinal}
\end{eqnarray}%
where $\Gamma =\left\vert \gamma _{1}\right\vert \nu _{F}$, $\mu (\mathbf{r}%
)=\mu -V_{\mathrm{trap}}(\mathbf{r})$, and

\begin{eqnarray}
&&K(\mathbf{r},\mathbf{r}^{\prime })=\sum_{n_{z}>0;n_{x},n_{y}\geq 0}\delta
_{n_{x}}\delta _{n_{y}}\ln [n_{z}+\frac{\omega _{\rho }}{\omega _{z}}%
(n_{x}+n_{y})]  \notag \\
&&\times \int_{M_{z}}^{1}d\zeta \int_{M_{x}}^{1}d\alpha
\int_{M_{y}}^{1}d\beta \ \delta (1-\zeta -\alpha -\beta )  \notag \\
&&\times \frac{4}{\pi ^{2}}\frac{\sqrt{(\zeta -z^{2})(\zeta -z^{\prime 2})}}{%
\zeta }\mathrm{U}_{n_{z}-1}(\frac{z}{\sqrt{\zeta }})\mathrm{U}_{n_{z}-1}(%
\frac{z^{\prime }}{\sqrt{\zeta }})  \notag \\
&&\times \frac{4}{\pi ^{2}}\frac{1}{\sqrt{(\alpha -x^{2})(\alpha -x^{\prime
2})}}\mathrm{T}_{n_{x}}(\frac{x}{\sqrt{\alpha }})\mathrm{T}_{n_{x}}(\frac{%
x^{\prime }}{\sqrt{\alpha }})  \notag \\
&&\times \frac{4}{\pi ^{2}}\frac{1}{\sqrt{(\beta -y^{2})(\beta -y^{\prime 2})%
}}\mathrm{T}_{n_{y}}(\frac{y}{\sqrt{\beta }})\mathrm{T}_{n_{y}}(\frac{%
y^{\prime }}{\sqrt{\beta }}),  \label{K}
\end{eqnarray}%
with $\delta _{n}=2$ for $n>0$, $\delta _{0}=1$, and $M_{s}=\max
(s^{2},s^{\prime 2})$ for $s=x,y,z$. The above equation can be viewed as the
equation for an extremum of the quadratic form

\begin{equation*}
F[\Delta _{0}]=\frac{1}{2}\int_{\mathbf{r},\mathbf{r}^{\prime }}\Delta _{0}(%
\mathbf{r})[L(\mathbf{r})\delta (\mathbf{r}-\mathbf{r}^{\prime })-K(\mathbf{r%
},\mathbf{r}^{\prime })]\Delta _{0}(\mathbf{r}^{\prime }),
\end{equation*}%
where $L(\mathbf{r})=3(1-r^{2})/\Gamma -(1-r^{2})^{3/2}\{\ln ([2\mu (\mathbf{%
r})/\omega _{z}]-2(4-\ln 4)/3\}$. The extremum can be found numerically
using the ansatz 
\begin{equation}
\Delta _{0}(\mathbf{r})=(1-r^{2})^{3/2}\sum_{m_{z},m_{\rho }}c_{m_{z}m_{\rho
}}\mathrm{U}_{m_{z}}(z^{2})\mathrm{T}_{m_{\rho }}(x^{2}+y^{2}),  \notag
\label{ansatz}
\end{equation}%
with $m_{z},m_{\rho }\geq 0$. The form $F$ becomes now a quadratic form of
the unknown coefficients $c_{m_{z}m_{\rho }}$, and the extremum corresponds
to the eigenvector of the matrix $M_{m_{z},m_{\rho },n_{z},n_{\rho }}$ of
the quadratic form with zero eigenvalue. Such an eigenvalue exists only if
the interaction $\Gamma $ and the trap frequencies $\omega _{i}$ obey
certain constraint. The parameter $\Gamma $ can be written as $\Gamma
=36n(0)d^{2}/\pi \mu $, where $n(0)=(2m\mu )^{3/2}/6\pi ^{2}\hbar ^{3}$ is
the gas density in the center of the trap, and hence, for a given dipole
moment $d$, depends only on the chemical potential $\mu $. On the other
hand, the chemical potential $\mu $ and the total number of particles $N$ in
the trap are related as $3N=\mu ^{3}/\omega _{z}\omega _{\rho }^{2}$.
Therefore, after fixing $\Gamma $ and $M$, the product of the trap
frequencies $\omega _{z}\omega _{\rho }^{2}$ is also fixed, and the only
free parameter left is the trap aspect ratio $\lambda $, which critical
value $\lambda _{c}$ could be determined from the above constraint.

The problem is therefore reduiced to finding the set of coefficients $%
c_{m_{z}m_{\rho }}$ and the aspect ratio $\lambda $ such that for a given $%
\Gamma $ and $N$ the lowest eigenvalue of the matrix $M_{m_{z},m_{\rho
},n_{z},n_{\rho }}$ of the quadratic form $F$ is zero. The matrix $M$ is
symmetric, and hence, there exists a solution.

The calculation of the matrix elements $M_{m_{z},m_{\rho },n_{z},n_{\rho }}$
is naturally divided into two parts (see Eq. (\ref{gapeqfinal})). The local
part with the kernel $L(\mathbf{r})$ is just a $3$-dimensional integral that
can be easily computed using, for instance, the Monte Carlo integration
routines, such as the VEGAS algorithm from the GSL library ~\cite{GSL}. The
non-local part with the kernel $K(\mathbf{r},\mathbf{r}^{\prime })$ is a
triple sum, which elements are $8$ dimensional integrals. The
straightforward approach using the same numerical method as described above
fails in this case, because it is too time consuming. To overcome this
problem, we perform the integrations over $\mathbf{r}$, $\mathbf{r}^{\prime
} $and $\zeta $ for fixed $\alpha $, $\beta $ and $n$'s using the same
method as before, and then use a $2$-dimensional spline interpolation method
to interpolate the integrand for the last integrations over $\alpha $ and $%
\beta $. In this way the time required to compute the matrix elements of $M$
for a given set of physical parameters, was reduced to about $72$ hours on a
standard workstation.

The results of the calculations are presented in 2 figures. Fig.1 shows the
desired relation betweem $\Gamma $ and the aspect ratio $\lambda $. The two
curves correspond to two different numbers of particles. As it could be
expected, for the larger number of particles, the critical aspect ratio $%
\lambda _{c}$ is smaller because the interaction is stronger. For a pancake
trap, $\lambda ^{-1}<1$, the interaction is predominantly repulsive, and
higher values of $\Gamma $ for fixed $\lambda $ are required to achieve the
BCS transition. On the other hand, for a cigar trap, $\lambda ^{-1}>1\,$,
the interaction is predominantly attractive and the occurrence of the BCS
transition requires smaller values of $\Gamma $.

Fig. 2 shows the order parameter $\Delta _{0}(\mathbf{r})$ for the cigar
trap with $\lambda ^{-1}=2.2$. Amazingly, the order parameter is a
nonmonotonic function of the distance from the trap center, in contrast to
the case of the BCS order parameter in a two component Fermi gas with short
range interactions ~\cite{monotonicDelta}. This effect persists, although
being less pronounced, for the case of a pancake geometry. In the axial
direction, the order parameter $\Delta (z,\rho =0)$ develops a minimum at $%
\rho <1$, whereas in the axial direction $\Delta (z=0,\rho )$ becomes
negative in the outer part of the cloud. This is a completely new behavior
originated from the anisotropy of the interpaticle interaction, that can
have profound consequences for the form and spectrum of the elementary
excitations. We expect an appearance of excitations with wave functions
concentrated in the inner region of the atomic cloud, where $\Delta \simeq 0$%
. This problem, however, goes beyond the scope of this paper.

Summarizing, we have determined the phase diagram for trapped dipolar Fermi
gases in the $\Gamma -\lambda^{-1}$ plane, where $\Gamma $ measures the
dipole strength and $\lambda $ is the trap aspect ratio. The BCS transition
at finite temperature $T$ is possible iff $\Gamma >\Gamma _{c}(\lambda )$.
We have calculated the critical line $\Gamma _{c}(\lambda )$, and the order
parameter at criticality.

We acknowledge support from the DFG, the RTN Cold Quantum Gases, ESF PESC
BEC2000+, Russian Foundation for Basic Research, and the Alexander von 
Humboldt Foundation.

\begin{figure}[tbp]
\epsfxsize7.0cm \centerline{
\epsfbox{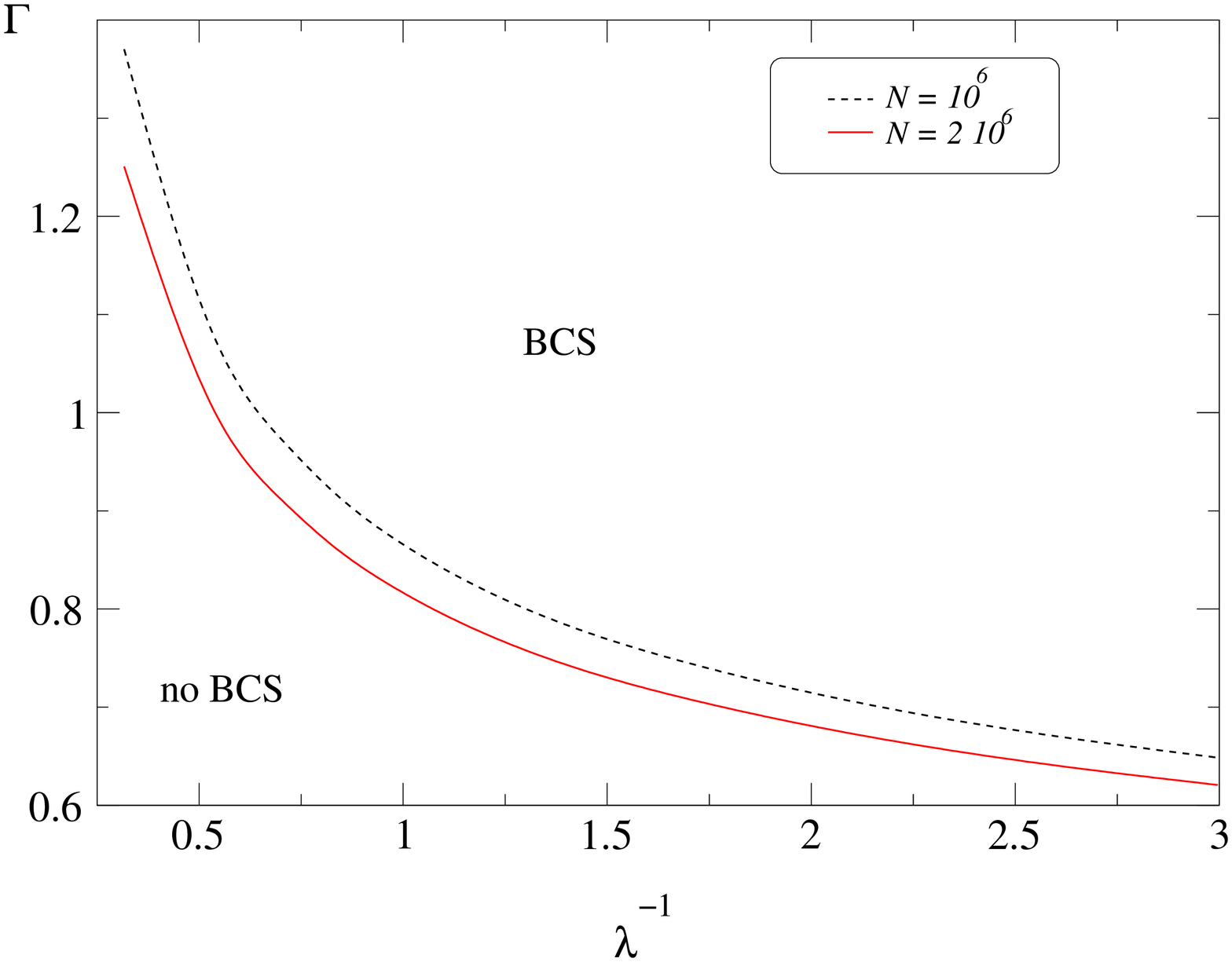}}
\caption{The critical $\Gamma _{c}$ as a function of the aspect ratio $%
\protect\lambda $: above the depicted curves BCS takes place. The upper
(lower) curve corresponds to $N=10^{6}$ and $N=2\times 10^{6}$.}
\label{fig:1}
\end{figure}

\begin{figure}[tbp]
\epsfxsize6.5cm \centerline{
\epsfbox{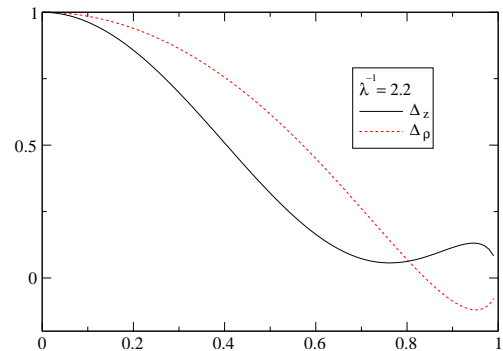}}
\caption{Order parameter for the aspect ratio $\protect\lambda ^{-1}=2.2$
(corresponding to cigar geometry). The solid line shows $\Delta _{0}(z,%
\protect\rho =0)$, and the dashed line corresponds to $\Delta _{0}({z=0,%
\protect\rho })$.}
\label{fig:4}
\end{figure}



\end{document}